\documentclass[runningheads]{llncs}
\usepackage[T1]{fontenc}
\usepackage{hyperref}

\usepackage{amssymb}
\usepackage{graphicx}
\usepackage{xcolor}
\graphicspath{ {./figures/} }
\usepackage{subcaption}
\usepackage{amsmath}

\newcommand{\myname}{PUMiNet}

\begin{document}
\title{PileUp Mitigation at the HL-LHC\\ Using Attention for Event-Wide Context\thanks{This work is supported by the U.S. Department of Energy (DoE) grant \emph{DE-SC0024669}}}
%
%

  \author{Luke Vaughan \textsuperscript{1} \and
  Mohammed Rakib \textsuperscript{2} \and
  Shivang Patel \textsuperscript{1} \and
  Flera Rizatdinova \textsuperscript{1} \and
  Alexander Khanov \textsuperscript{1} \and
  Arunkumar Bagavathi \textsuperscript{2}
  }

 \authorrunning{L. Vaughan, M. Rakib, S. Patel, F. Rizatdinova, A. Khanov, A. Bagavathi}

  \institute{\textsuperscript{1} Department of Physics, \textsuperscript{2} Department of Computer Science \\
  Oklahoma State University \\
  \email{\{luke.vaughan, mohammed.rakib, shivang.patel, flera.rizatdinova, alexander.khanov, abagava\}@okstate.edu}}

\maketitle              
\begin{abstract}
The Large Hadron Collider, LHC, collides bunches of protons resulting in multiple interactions that occur practically simultaneously. This creates a pileup effect that distorts physics measurements due to the products of pileup collisions. In order to improve the discovery potential of the LHC, it is necessary to mitigate the effect of pileup interactions on the processes of interest. In this paper, we suggest a novel AI-based method, \myname{}, to tackle the problem of pileup at the current LHC and future High Luminosity LHC conditions. \myname{} is an attention-based algorithm that mitigates pileup effects using a regression task on jets in the context of an entire event. At $\left\langle \mu \right\rangle=200$, \myname{} is able to predict the hard scatter energy and mass fractions of jets with $R^2=0.912$ and $R^2=0.720$, respectively. These predictions enable the reconstruction of the Higgs boson mass in the HL-LHC environment.

\end{abstract}
\section{Introduction}\hfill

At the Large Hadron Collider, LHC, the ATLAS and CMS experiments study the interactions of elementary particles to better understand the fundamental structure of nature. Experiments are performed by colliding groups of protons, or bunches, at extremely high energies. At such high energy, each collision leads to the production of hundreds of particles which are registered by the ATLAS and CMS detectors. The information recorded by the detectors are referred to as an \emph{event}. Since each bunch contains $10^{11}$ protons, each event contains information from multiple proton-proton interactions. The average number of interactions per bunch crossing recorded in an event is referred to as $\left\langle \mu \right\rangle$. The events selected for recording are chosen in such a way that at least one of these interactions is a \emph{hard scatter}, considered as \emph{signal}, which contains highly energetic particles that are of interest to physics. Other interactions recorded in the same event are called \emph{pileup}, considered as \emph{background}, which contaminates the signal.

\begin{figure}[ht]
\centering
  \includegraphics[width=0.8\linewidth]{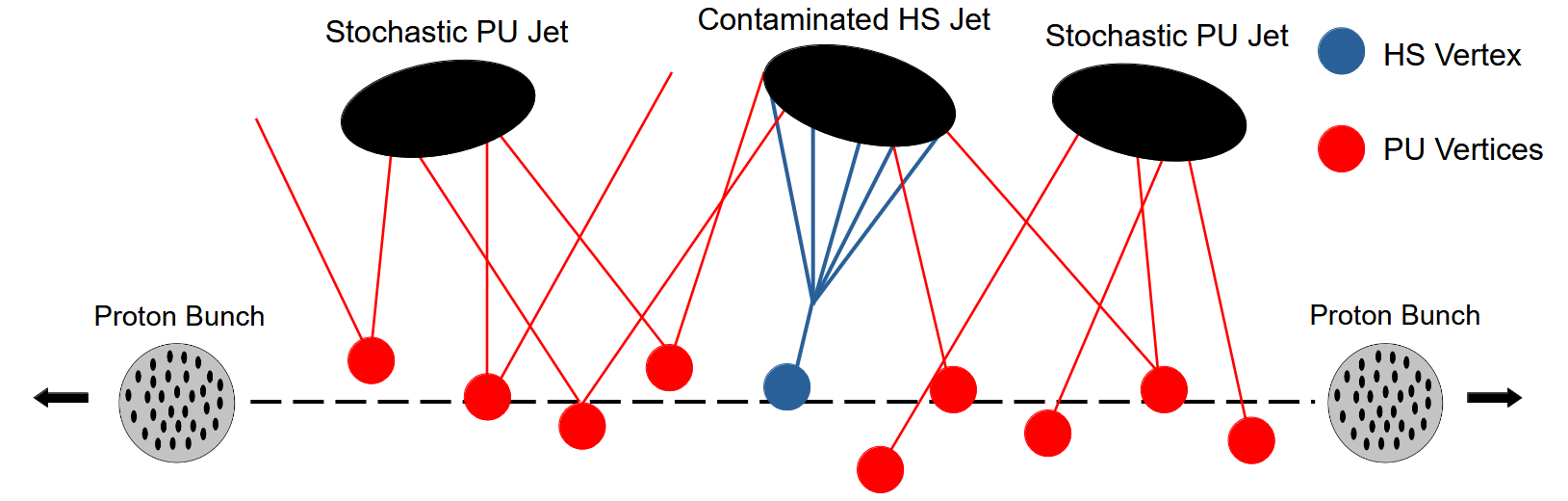} 
  \caption{The interactions from crossing proton bunches in a single event. The hard scatter, HS, originates from the primary vertex forming a correlated set of particles while other interactions stochastically produce pileup, PU.}
\label{fig:PileupJets}
\end{figure}

Our research focuses on physics processes where particles form streams that can be clustered into tightly knit cones called \emph{jets}, which can originate from both hard scatter and pileup collisions. Throughout this article, we refer to particles as \emph{tracks}, and the collision points from which the tracks originate as \emph{vertices}. In a real experiment, jets are measured and identified using information from many subdetector systems, but to simplify the problem, this paper treats jets purely as sets of tracks. Jets containing high fraction of hard scatter tracks are of actual physics interest, but there often exists many other jets originating from the pileup interactions, shown in Figure~\ref{fig:PileupJets}.

The LHC prepares to upgrade to the High-Luminosity phase, HL-LHC, where $\left\langle \mu \right\rangle$ will increase from 60 to 200. This upgrade will dramatically increase the number of pileup particles in each event, which will increase the statistics of rare hard scatter processes. However, in order to maximize the discovery potential of the HL-LHC, there is a need to mitigate high pileup conditions: in particular, to quantify the energy and momentum contributions from pileup to hard scatter jets. Existing pileup mitigation techniques have considered two distinct approaches: (i) jet-level pileup mitigation~\cite{ATLAS-CONF-2014-018}, or (ii) track-level pileup mitigation~\cite{Bertolini_2014}. The authors choose to focus on jet-level pileup mitigation, and therefore models implementing track-level pileup mitigation will be considered out of scope for the benchmarking procedure.

Machine learning methods have become widespread in High Energy Physics, HEP, ~\cite{he2023high,Larkoski_2020} assisting with pileup mitigation~\cite{komiske2017pileup,ABCNet,Maier_2022}, jet tagging~\cite{ParticleNet}, top tagging~\cite{Barman_2024}, event modeling~\cite{kansal2021particle} and searches for new physics using anomaly detection~\cite{duarte2024novelmachinelearningapplications} and other AI methods. The ATLAS experiment is using the JVT kNN algorithm~\cite{ATLAS-CONF-2014-018} which relies on constructing high level variables using track and jet features to perform binary classification on a per jet basis. However, set-based neural networks ~\cite{zaheer2018deepsets} and message passing graph neural networks can further exploit the set structure of jets and have seen great success in jet classification tasks~\cite{ParticleNet}. Furthermore, it has been shown~\cite{qu2022particle} that for the task of jet classification, attention neural networks can achieve higher accuracy with fewer number of parameters. Therefore, attention neural networks are suitable for high pileup conditions due to their strengths representing set-based data structures and their ability to scale more efficiently than GNN. Attention can efficiently capture context from an entire event which was previously considered computationally unfeasible for GNNs. 

Therefore, this work proposes an attention-based neural network, \myname{}, to mitigate pileup by performing the simultaneous identification, energy correction, and mass correction of hard scatter jets. \myname{} uses self and cross attention to capture all possible correlations between jets and tracks within an entire event while maintaining good performance when scaling to high pileup conditions. \myname{} also outperforms benchmarks that consider jet-level pileup mitigation and increases the discovery potential of the HL-LHC through a practical example of a physics analysis.

\section{Jet Label Definitions}\label{JetLabels}\hfill

Both jets and tracks are characterized by Lorentz 4-vectors $\overrightarrow{J}$ and $\overrightarrow{T}$, respectively. A Lorentz 4-vector is defined as $(E,\overrightarrow{p})$, where E is energy and $\overrightarrow{p}=(p_x,p_y,p_z)$ is the momentum in 3D space. The mass of an object described by a Lorentz 4-vector is defined as $m=\sqrt{E^2-|\overrightarrow{p}|^2}$.\footnote{In Natural Units} Each event, $\mathcal{E}$, is composed of a variable number of jets and each jet is formed by a variable number of associated tracks. Therefore, the fundamental data structure of an event, $\mathcal{E}$, can be interpreted as nested sets. The relationship between jet 4-vector $\overrightarrow{J}$ and the associated track 4-vectors, $\overrightarrow{T}$, is:

\begin{equation}
  \overrightarrow{J}_i = \sum\limits_{j \in \overrightarrow{J}_i} \overrightarrow{T}_{ij}
\end{equation}

Most of the time, jets have contributions from both hard scatter and pileup, and the goal is to purely recover the hard scatter component. However, the actual measured quantities of total energy, $E_{jet}$, and total mass, $m_{jet}$, of each jet contain pileup contamination. To mitigate the effects of pileup, we propose the use of continuous fractions, $E_{frac}$ and $M_{frac}$,\footnote{$E_{frac}$ and $M_{frac}$ can be considered scalar and vector corrections to $\overrightarrow{J}$, respectively. Pileup's stochastic directions affects the vector corrections significantly more than scalar corrections.} to be directly applied to each jet, as ratios, in order to recover only the hard scatter energy, $E_{HS}$, and mass, $m_{HS}$. Therefore, truth labels can be constructed from the 4-vectors of each jet, $\overrightarrow{J}$, by defining:
\begin{equation}
  E_{frac}=\frac{E_{HS}}{E_{jet}} \phantom{.......} M_{frac}=\frac{m_{HS}}{m_{jet}}.
\end{equation}


For the first time, this article proposes to simplify the pileup mitigation strategy by directly applying $E_{frac}$ and $M_{frac}$ labels to simultaneously identify pileup and apply corrections to hard scatter jets through the use of attention neural networks. This approach uses all available information from an event, including the correlations manifesting in hard scatter processes, to get a more accurate representation of the underlying physics process. Using a simulated dataset, the proposed approach outperforms traditional pileup mitigation strategies and assists with physics analysis.

\section{Simulated Dataset}\label{Dataset}\hfill

Jets and tracks measured by the ATLAS and CMS detectors are described using a 3D detector coordinate system with their transverse momentum, $p_{\rm T}$, and angular components, $\eta$, and $\phi$.\footnote{$\phi$ is azimuthal angle, $\eta=-\ln\left[ \tan\left( \frac{\theta}{2} \right) \right]$ where $\theta$ is polar angle, and $p_{\rm T}=|\overrightarrow{p}|\sin\theta$ in the spherical coordinate system where the $z$ axis is directed along the beam.}
As $\left\langle \mu \right\rangle$ increases, the pileup tracks form more jets, leading to a much higher number of jets per event at the same $p_{\rm T}$. Jet features include: $[p_{\rm T},\eta,\phi,m]$, and track features include: $[p_{\rm T},\eta,\phi,m,d_0,z_0]$, where $d_0$ and $z_0$ are transverse and longitudinal distance to primary vertex when track is extrapolated to beam line using parameterized estimation described in~\cite{ATLAS:2021yvc}.


To evaluate the performance of \myname{}, a specific physics process is chosen: the simultaneous production of two Higgs bosons decaying into a $b$-anti $b$ quark pair. This process is of high interest for HEP and is one of the key components for the HL-LHC physics program~\cite{Dainese:2019rgk}. Each Higgs boson gives rise to a pair of jets $\overrightarrow{J}_i$, $\overrightarrow{J}_j$, described by a combined 4-vector $\overrightarrow{J}_{ij}$. When paired through the proper combinatorics, the distribution of the $\overrightarrow{J}_{ij}$ mass forms a resonance peak near the expected Higgs mass, $m_{\rm H}\approx125$~GeV, but pileup contamination will shift and widen the peak.


A sample of 50k di-Higgs events is generated using MadGraph interfaced with Pythia \footnote{The process \texttt{p p > h h, h > b b\~{}} generated with MadGraph5\_aMC@NLO v3.5.5~\cite{madgraph} interfaced to Pythia 8.312~\cite{pythia}.}. Samples are simulated at $\left\langle \mu \right\rangle = 60$ for the LHC pileup conditions and at $\left\langle \mu \right\rangle = 200$ for the HL-LHC pileup conditions. Pileup processes\footnote{The process \texttt{SoftQCD:inelastic} generated with Pythia using the A14 central tune~\cite{TheATLAScollaboration:2014rfk} and NNPDF2.3LO~\cite{BALL2013244}.} are overlaid according to a Poisson distribution with mean of $\left<\mu\right>$. The position of each hard scatter and pileup vertex in the event is independently smeared according to a Gaussian distribution with $\sigma_{x,y}=0.3$~mm in transverse plane and $\sigma_z=50$~mm along the direction of the proton beam to mimic the actual LHC conditions. Stable particles are then passed to FastJet~\cite{Cacciari_2012} to be clustered into jets using anti-$k_t$ algorithm~\cite{Cacciari_2008} with the cone size $R=0.4$ \footnote{$R = \sqrt{( \eta_{jet} - \eta_{track})^2 + ( \phi_{jet} - \phi_{track})^2}$, where the $\phi$ difference is taken modulo $2\pi$.} and minimum jet $p_{\rm T}$ of 25~GeV. To incorporate detector limitations, neutral particles and particles with $p_{\rm T}$ below 400~MeV are removed from the dataset.

\section{Methodology}\hfill

\begin{figure}[t]
\centering
  \includegraphics[width=1\linewidth]{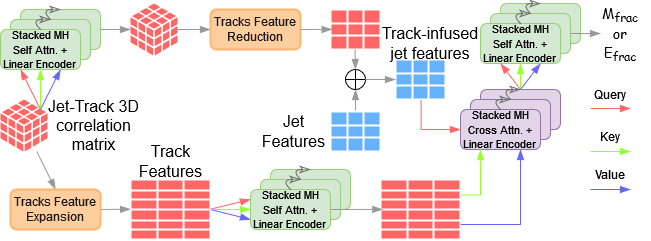}
\caption{Architecture of the proposed attention-based neural network method. Our method extracts two versions of track features to combine with jet features. The proposed multi-head cross-attention block correlates jets with respect to all tracks to enable learning of jet features based on an entire event.}
\label{fig:Model}
\end{figure}

We introduce a machine learning approach \myname{} that incorporates jets and tracks to quantify $E_{frac}$ and $M_{frac}$ of jets. We define this problem as a regression task: $f(\mathcal{J},\mathcal{T})[\theta] \Rightarrow M_{frac}$ and $g(\mathcal{J},\mathcal{T})[\psi] \Rightarrow E_{frac}$, where $\theta$ and $\psi$ are model parameters. We consider $\mathcal{J} \in \mathbb{R}^{N_J \times x}$, and $\mathcal{T} \in \mathbb{R}^{N_J \times N_T \times y}$, where $x$ is jet feature dimensions, $y$ is track feature dimensions, $N_J$ is the total number of jets in an event, $N_T$ is the total number of tracks in a given jet. Our aim in this work is to extract rich contextual track information that exists in the jet-track correlation matrix $\mathcal{T}$ to reinforce jet features for the underlying jet-level regression task. We depict the overall architecture of the proposed model in Figure~\ref{fig:Model}.

\textbf{Transformer Encoders}: We incorporate encoders with two forms of Set Attention ~\cite{SetAttention} in this work: (i) \emph{Encoders with self-attention} to capture the dependencies of jets and tracks separately, and (ii) \emph{Encoder with cross-attention} to capture the dependencies between jets and tracks on each other. We employ the transformer encoder, which operates on inputs \(Q\) (query), \(K\) (key), and \(V\) (value). First, we apply scaled-dot product attention on normalized vectors to calculate the attention weights and extract correlations: $\text{Attention}(\mathbf{Q}, \mathbf{K}, \mathbf{V}) = \text{softmax}\left(\frac{\mathbf{Q} \mathbf{K}^\top}{\sqrt{d_E}}\right) \mathbf{V}$. We use multi-head attention, $MHA(Q,K,V) = $\\
$ Concat(Attention_1,\ldots,Attention_h)$, where $h$ is the total number of attention heads in the encoder. Second, a residual connections is used to generate the context vector: $\mathbf{Q}_\text{Context} = \mathbf{Q} + \text{MHA}(\mathbf{Q}, \mathbf{K}, \mathbf{V})$. Third, a feed-forward network and an additional skip connection are used to produce the updated representation of the input query: $\mathbf{Q}_\text{Output} = \mathbf{Q}_{\text{Context}} + \text{FFN}(\mathbf{Q}_{\text{Context}})$. This encoder block can be stacked numerous times to iteratively update the query with various key-value pairs. Therefore, the two types of encoders in \myname{} can be represented as given in Equation~\ref{eq:encoders} for inputs $\mathbb{X}$ and $\mathbb{Y}$:

\begin{equation}
    \text{Self-Encoder}(\mathbb{X},\mathbb{X},\mathbb{X}) = \mathbb{X}' \quad \text{Cross-Encoder}(\mathbb{X},\mathbb{Y},\mathbb{Y}) = \mathbb{X}'
    \label{eq:encoders}
\end{equation}

\textbf{Learning Local Track-reinforced Jet Features $\hat{\mathcal{J}}$}: Apart from using jet features $\mathcal{J}$ for the underlying regression problems $f(\theta)$ and $g(\psi)$, transformer encoders can enrich a jet with its associated tracks for a more complete jet representation. To aggregate track features in jets, we first correlate jet and track features in the jet-track matrix $\mathcal{T}$ using $\mathcal{T}^{\prime} = \text{Self-Encoder}(\mathcal{T},\mathcal{T},\mathcal{T})$. We then aggregate tracks features of jets $\mathcal{J}_i$ using the sum operation, as given in Equation~\ref{eq:reduction}, where $N_T(\mathcal{J}_i)$ is the number of tracks in the jet $\mathcal{J}_i$. We then align the aggregated jet features $\hat{\mathcal{T}} \in \mathbb{R}^{N_J \times y}$ with the real jet features using the concatenation operation and apply non-linear transformation: $\hat{\mathcal{J}} = \text{ReLU}([\mathcal{J},\hat{\mathcal{T}}]W + b)$, where $W$ and $b$ are trainable weights and biases respectively.

\begin{equation}
\hat{\mathcal{T}}(\mathcal{J}_i) = \sum_{t=1}^{N_T(\mathcal{J}_i)} \mathcal{T}_{t}^{\prime}
\label{eq:reduction}
\end{equation}

\textbf{Learning Global Track Features $t$}: Apart from reinforcing jets with correlations based on local track features, we also aim to enrich jet features independently based on global track correlations during each event. To this end, we first flatten the jet-track correlation matrix $\mathcal{T}$ to consider only the track features $t \in \mathbb{R}^{N_{T} \ \times y}$, where $N_{T}$ is the total number of tracks in the event. We then enrich track features using the stacked self-encoder module: $\hat{t} = \text{Self-Encoder}(t,t,t)$. We emphasize that self-encoder modules on all tracks capture global contextual details in the event. 

\textbf{Learning Global Jet Features}: To model diverse event-wide correlations between jets $\hat{\mathcal{J}}$ and independent tracks $\hat{t}$, we apply encoder with cross-attention as $\mathcal{J}_{\text{cross}} = \text{Cross-Encoder}(\hat{\mathcal{J}},\hat{t},\hat{t})$. These encoded jet representations are based on the global context given by independent track representations that are present both within and outside the jets. Such unique representations allow the model to capture all possible correlations present in hard-scatter processes and give importance to learning jet features at the event-level. We further enrich these jet representations using $\mathcal{J}_{\text{final}} = \text{Self-Encoder}(\mathcal{J}_{\text{cross}},\mathcal{J}_{\text{cross}},\mathcal{J}_{\text{cross}})$.

\textbf{Learning Objective}: We predict $E_{frac}$ and $M_{frac}$ using final jet features \(\mathcal{J}_{\text{final}}\) by attaching the regression layer with sigmoid activation function ($\sigma$): $\hat{y} = \sigma(\mathcal{J}_{\text{final}} \mathbf{W}_{\text{out}} + \mathbf{b}_{\text{out}})$, where $\mathbf{W}_{\text{out}}$ and $\mathbf{b}_{\text{out}})$ are learnable regression weights and biases. We optimize all model parameters to learn optimal jet and track features in a supervised fashion using the Mean Squared Error loss function:
\[
\mathcal{L} = \frac{1}{N} \sum_{i=1}^N \left( \hat{y}_i - y_i \right)^2,
\]
where \(N\) is the total number of training samples, $y$ is the ground truth ($E_{frac}$ or $M_{frac}$), and $\hat{y}$ is the model prediction. By leveraging track features and global context through iterative encoder-based attention mechanisms, the model progressively refines jet features to predict energy and mass fractions with least mean square error.

\section{Experiments and Results}\hfill

\textbf{\myname{} Setup:} We deploy \myname{} with each encoder block stacked three times to achieve a deeper representation of jets. We use embedding dimension of 256 and 8 attention heads per encoder. We schedule the learning rate to decay by a factor of 0.1 after the $25^{th}$ epoch which noticeably helped descend the noisy loss landscape in model training. The \emph{AdamW} optimizer is used to prevent overtraining using weight decay and we use $R^2$ as an evaluation metric. The model converged after training 50 epochs on an NVIDIA RTX 3090.\\

\begin{figure}[ht!]
\begin{subfigure}{.25\textwidth}
  \includegraphics[width=1\linewidth]{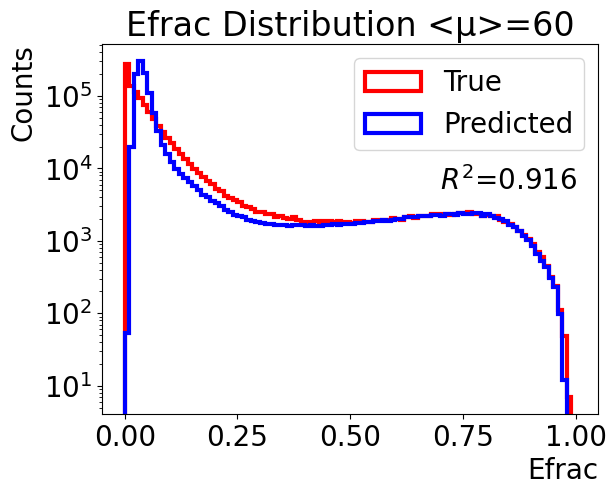}
  \caption{}
  \label{fig:Efrac1d_mu60}
\end{subfigure}%
\begin{subfigure}{.25\textwidth}
  \includegraphics[width=1\linewidth]{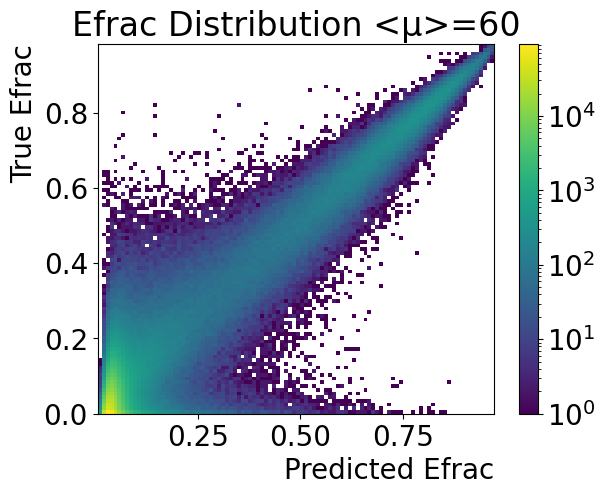}
  \caption{}
  \label{fig:Efrac2d_mu60}
\end{subfigure}
\begin{subfigure}{.25\textwidth}
  \includegraphics[width=1\linewidth]{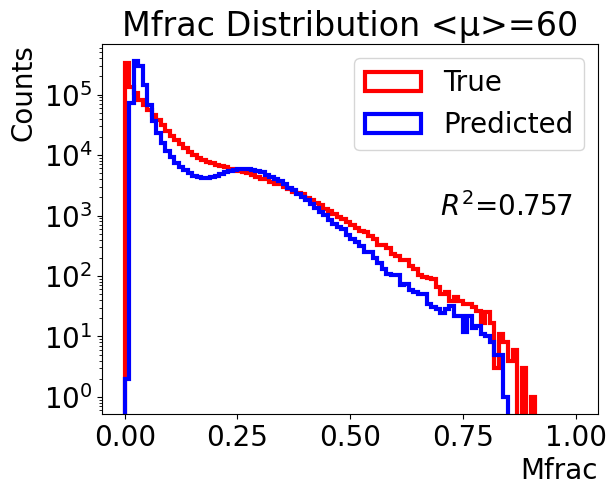}
  \caption{}
  \label{fig:Mfrac1d_mu60}
\end{subfigure}%
\begin{subfigure}{.25\textwidth}
  \includegraphics[width=1\linewidth]{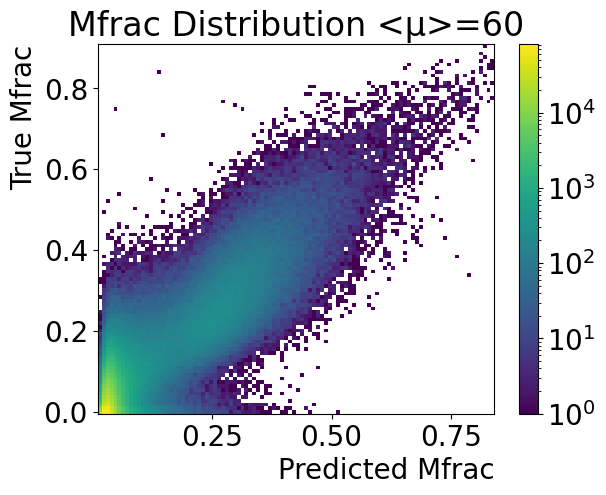}
  \caption{}
  \label{fig:Mfrac2d_mu60}
\end{subfigure}
\begin{subfigure}{.25\textwidth}
  \includegraphics[width=1\linewidth]{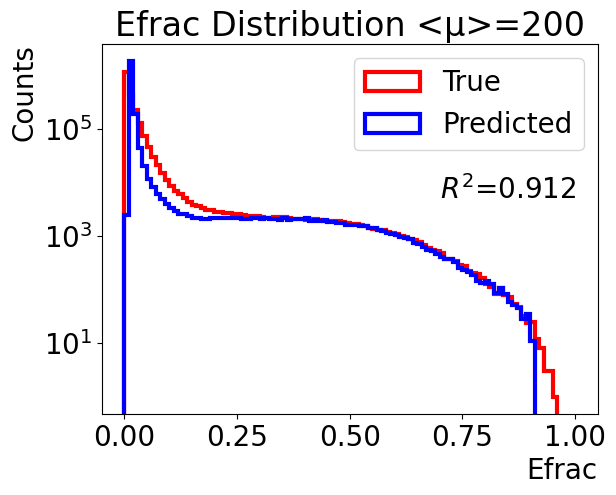}
  \caption{}
  \label{fig:Efrac1d_mu200}
\end{subfigure}%
\begin{subfigure}{.25\textwidth}
  \includegraphics[width=1\linewidth]{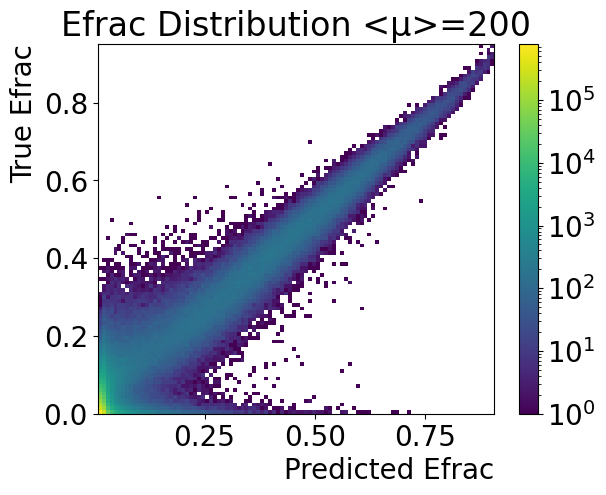}
  \caption{}
  \label{fig:Efrac2d_mu200}
\end{subfigure}
\begin{subfigure}{.25\textwidth}
  \includegraphics[width=1\linewidth]{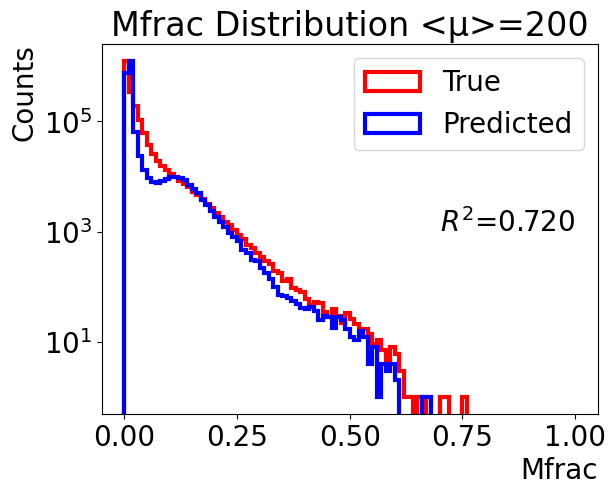}
  \caption{}
  \label{fig:Mfrac1d_mu200}
\end{subfigure}%
\begin{subfigure}{.25\textwidth}
  \includegraphics[width=1\linewidth]{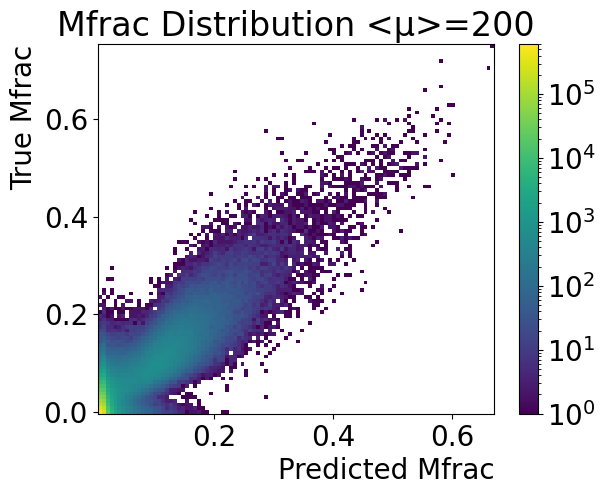}
  \caption{}
  \label{fig:Mfrac2d_mu200}
\end{subfigure}
\caption{At $\left \langle \mu \right \rangle=60$ (top row) and $\left \langle \mu \right \rangle=200$ (bottom row), the predicted energy (left) and mass (right) fraction of jets shown as 1D and 2D histograms.}
\label{fig:RegressionResults}
\end{figure}

\textbf{Regression Results:} \myname{} was evaluated on a sample of 20k di-Higgs events simulated specifically for testing. At $\left \langle \mu \right \rangle=60$, the model achieves $R^2=0.916$ for $E_{frac}$, as shown in Figure~\ref{fig:Efrac1d_mu60}, and $R^2=0.757$ for $M_{frac}$, as shown in Figure~\ref{fig:Mfrac1d_mu60}. At $\left \langle \mu \right \rangle=200$, the model achieves $R^2=0.912$ for $E_{frac}$, as shown in Figure \ref{fig:Efrac1d_mu200}, and $R^2=0.720$ for $M_{frac}$, as shown in Figure \ref{fig:Mfrac1d_mu200}. The 2D predicted vs truth values, plotted with a log z color scale, shown in Figure [\ref{fig:Efrac2d_mu60},\ref{fig:Mfrac2d_mu60}] at $\left \langle \mu \right \rangle=60$ and in Figure \ref{fig:Efrac2d_mu200} \ref{fig:Mfrac2d_mu200} at $\left \langle \mu \right \rangle=200$, show that there is good diagonal trend between the predictions and the truth. Overall, the transformer encoder architecture provides a highly parallelizable algorithm that is computationally feasible at high pileup conditions, and the plots in Figure~\ref{fig:RegressionResults} show that \myname{} learns the hard scatter contributions without significant degradation to the $R^2$ value at high pileup conditions of the HL-LHC.

\newpage
\textbf{Classic JVT Benchmark:} In order to benchmark the performance of the model against existing JVT algorithm, only a subset of the model is used to maintain consistency across input features. For a fair comparison, only jet features are considered: \{$p_T,\eta,\phi,m,R_{p_T},corrJVF$\} where $R_{p_T}$ and $corrJVT$ are defined in ~\cite{ATLAS-CONF-2014-018}. This exercise attempts to show that the attention architecture (AttnJVT) alone brings noticeable improvements when jets are processed in the context of an event. These results can be further improved when jets are compounded with track features.

\begin{figure}[ht]
\centering
\begin{subfigure}{.32\textwidth}
  \centering
  \includegraphics[width=1\linewidth]{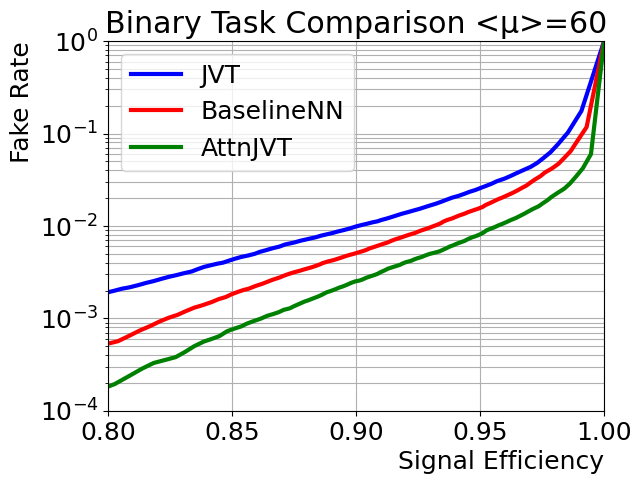}
  \caption{}
  \label{fig:Benchmark:sub1}
\end{subfigure}%
\begin{subfigure}{.32\textwidth}
  \centering
  \includegraphics[width=1\linewidth]{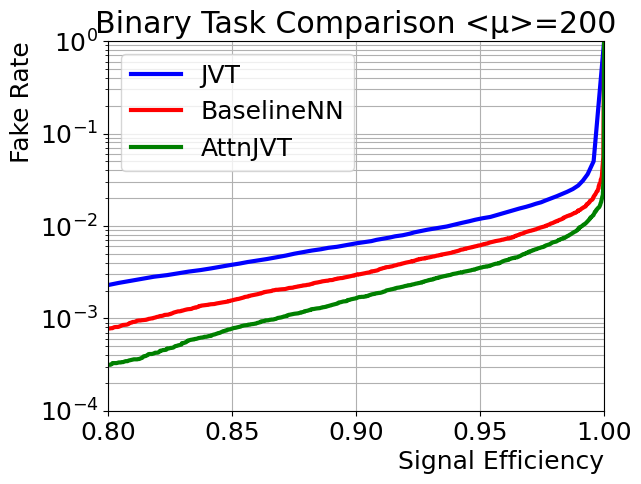}
  \caption{}
  \label{fig:Benchmark:sub2}
\end{subfigure}
\begin{subfigure}{.32\textwidth}
  \centering
  \includegraphics[width=1\linewidth]{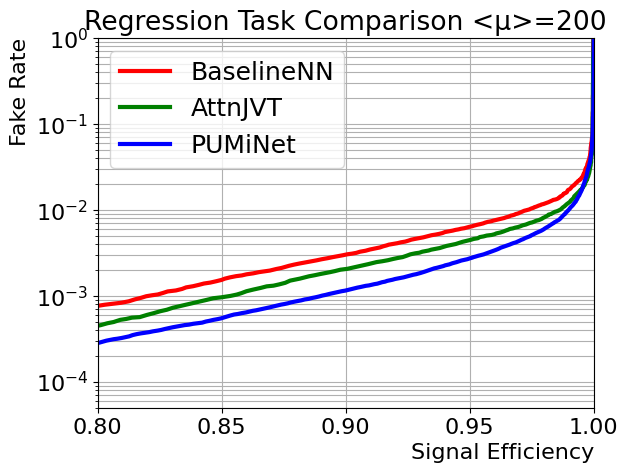}
  \caption{}
  \label{fig:Benchmark:sub3}
\end{subfigure}
\caption{Benchmark performance for the binary classification task for replica ATLAS JVT (blue), baseline deep NN (red), and MHA jet encoder (green) for for $\left<\mu\right>=60$ (a) and $\left<\mu\right>=200$ (b). Further improvement can be gained by \myname{} when tracks are added to the model (c). }
\label{fig:Benchmark}
\end{figure}

Using the di-Higgs dataset, the following models were created: (1) a replica of ATLAS JVT kNN model described in ~\cite{ATLAS-CONF-2014-018} (2) a baseline deep neural network, and (3) a single MHA encoder between jets, AttnJVT.\footnote{A small number of learnable parameters are used, 21k, for AttnJVT and the baseline deep NN.} Since JVT is a binary classifier, the continuous labels are converted to binary using a cut at $E_{frac}=0.3$. The JVT algorithm and deep NN process inputs on a per jet basis, but AttnJVT processes an entire set of jets in the context of an event which allows the model to capture correlations between hard scatter jets. The baseline deep NN and AttnJVT are trained using the Binary Cross Entropy loss function, and converges after 60 epochs on 50k events. The benchmarked false positive rates against the true positive rates are shown in Figure Figure \ref{fig:Benchmark:sub1} \& \ref{fig:Benchmark:sub2} which shows that AttnJVT can significantly lower the false positive rate. We also show that \myname{} brings further improvements over AttnJVT using both jet and track features to capture event-wide correlations in the regression setup, as shown in Figure \ref{fig:Benchmark:sub3}.

\begin{table}[htbp]
 \begin{center}
  \caption{Hyperparameter Scan at $\left\langle \mu \right\rangle = 60$}
  \begin{tabular}{|c|c|c|c|}
  \hline
    \hline
    Embed Dim & Attn Heads & $E_{frac}$ $R^2$ & $M_{frac}$ $R^2$ \\
    \hline
    \hline
    64   & 8  & 0.909 & 0.748 \\
    \hline
    128   & 8  & 0.919 & 0.755 \\
    \hline
    256   & 8  & 0.916 & 0.757 \\
    \hline
    256   & 16  & 0.918 & 0.761 \\
    \hline
    256   & 32  & 0.916 & 0.757 \\
    \hline
  \end{tabular}
 \label{tab:Albation}
 \end{center}
\end{table}

Lastly, to determine the robustness of \myname{}, a parameter scan was performed by changing the embedding dimension and number of attention heads. The results of this study are summarized in Table \ref{tab:Albation}. In general, we believe the results are within expected uncertainties and are evidence of a stable minimum with higher embedding dimension performing slightly better.

\section{Physics Case Study}\hfill

To demonstrate that the model is able to provide useful insight into a practical physics analysis, we attempt to reconstruct the Higgs boson mass from the simulated di-Higgs dataset along with a non-resonant multijet background sample (denoted 4b). 

\begin{figure}[ht]
\centering
\begin{subfigure}{.32\textwidth}
  \centering
  \textbf{\tiny{Raw Mass $\left \langle \mu \right \rangle=200$}}
  \includegraphics[width=1\linewidth]{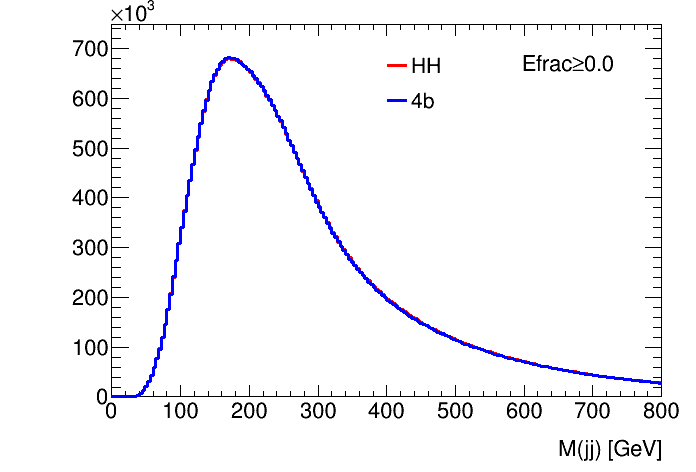}
  \caption{}
  \label{fig:Raw}
\end{subfigure}
\begin{subfigure}{.32\textwidth}
  \centering
  \textbf{\tiny{Uncorrected Mass $\left \langle \mu \right \rangle=200$}}
  \includegraphics[width=1\linewidth]{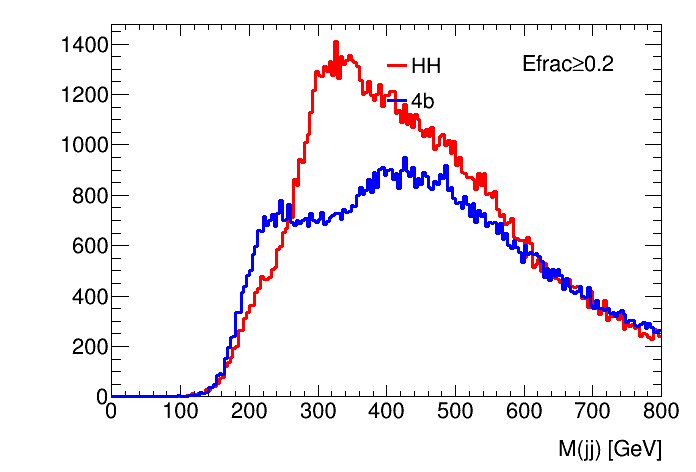}
  \caption{}
  \label{fig:Uncorrected}
\end{subfigure}
\begin{subfigure}{.32\textwidth}
  \centering
  \textbf{\tiny{Corrected Mass $\left \langle \mu \right \rangle=200$}}
  \includegraphics[width=1\linewidth]{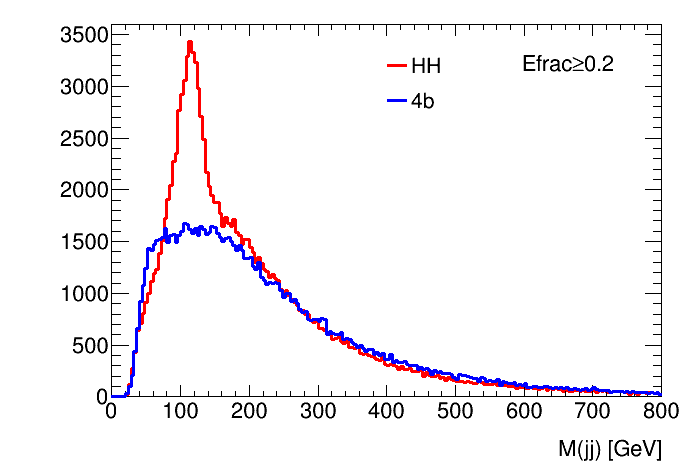}
  \caption{}
  \label{fig:Corrected}
\end{subfigure}
\caption{The reconstructed Higgs boson mass when looking at (a) all jets with no cuts on $E_{frac}$ and no corrections, (b) uncorrected jets with cut at true $E_{frac}>0.2$, and (c) corrected jets (according to model predictions) with cut at predicted $E_{frac}>0.2$. (a) Shows no signs of mass peak due to pileup contamination. (b) Shows a mass peak that is heavily smeared due to pileup. (c) Shows the expected narrow narrow peak near $m_{\rm H}\approx 125$~GeV with corrections applied.}
\label{fig:MassPeak}
\end{figure}

The background sample was generated as the MadGraph process\newline\texttt{p p > b b\~{} b b\~{}} with the minimum $p_{\rm T}$ of the $b$-quarks set to 60~GeV to ensure the same kinematics of both samples, so that the only visible difference between the samples is the resonant mass peak that appears in the di-Higgs sample. Figure \ref{fig:MassPeak} shows the reconstructed mass of all possible pairs of jets in each event, and we expect to find a resonance mass peak near $m_{\rm H}\approx 125$~GeV with a narrow width. Figure \ref{fig:Raw} shows the uncorrected jets with no cut applied to $E_{frac}$ and the background and signal are indistinguishable. Figure \ref{fig:Uncorrected} shows that a cut at truth level $E_{frac}>0.2$ shows a resonant mass peak with uncorrected jets, however, the mean and the width of the mass peak has been significantly inflated due to the effects of pileup. Figure \ref{fig:Corrected} shows that a cut a prediction level $E_{frac}>0.2$ on corrected jets shows a mass peak near the expected value of 125~GeV with a narrow width which shows that the use of $E_{frac}$ and $M_{frac}$ to correct jets successfully mitigates the effects of pileup and restores physical quantities to their expected values.

\begin{figure}[h]
\centering
\begin{subfigure}{.55\textwidth}
  \includegraphics[width=1\linewidth]{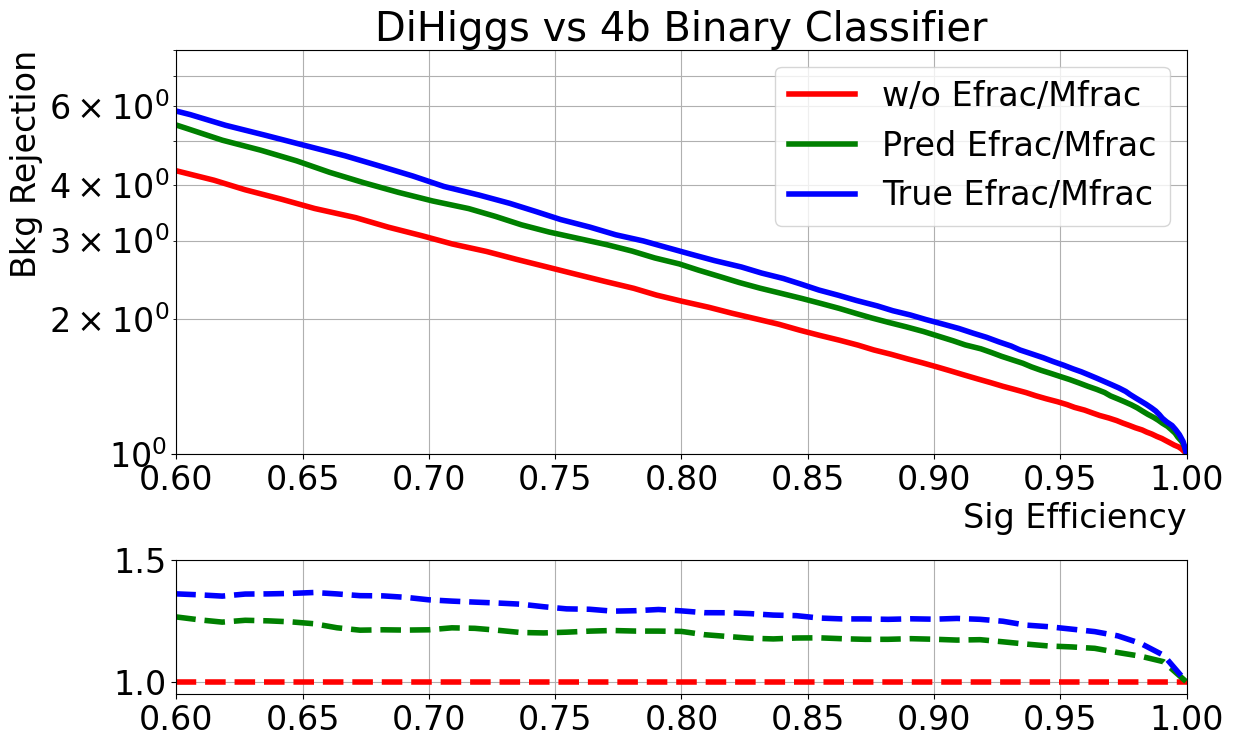}
  \caption{}
  \label{fig:ROC}
\end{subfigure}%
\begin{subfigure}{.43\textwidth}
  \includegraphics[width=1\linewidth]{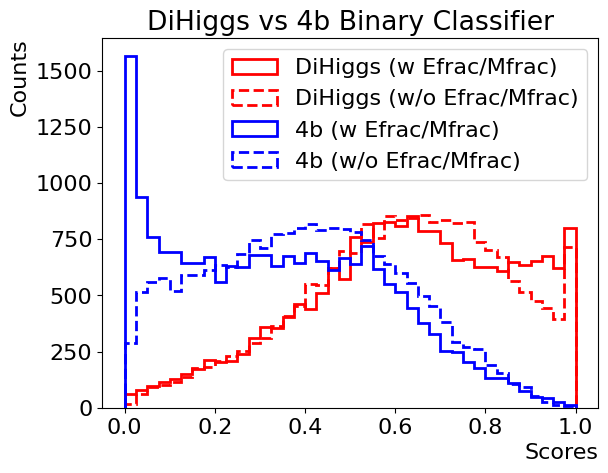}
  \caption{}
  \label{fig:Scores}
\end{subfigure}
\caption{For the purposes of physics analysis, the learning objective is modified to perform direct binary classification of di-Higgs vs 4b. $E_{frac}$ and $M_{frac}$ improve performance of the classifier.}
\label{fig:Analysis}
\end{figure}

Lastly, we would like to show that one can slightly modify the learning objective of the model to directly classify an event as signal or background through a binary classification task, which circumvents the need for computationally costly combinatorics. To demonstrate this effect, a self-encoder module between jets is trained on a binary classification task, di-Higgs vs 4b, for three different cases where jets are represented by the following features: (1) $[p_{\rm T},\eta,\phi,m]$, (2) $[p_{\rm T},\eta,\phi,m,E_{frac}^{pred}, M_{frac}^{pred}]$, and (3) $[p_{\rm T},\eta,\phi,m,E_{frac}^{true}, M_{frac}^{true}]$. Through learning the proper attention weights between jets, the model is able to perceive the mass peak and directly classify the event as di-Higgs signal or 4b background. Figure \ref{fig:ROC} shows the background rejection, the reciprocal of false positive rate, vs the signal efficiency, the true positive rate. 

For the case (1) \myname{} is able to successfully distinguish between di-Higgs and 4b physics processes despite having quite similar kinematic features. When predicted and truth $E_{frac}$ and $M_{frac}$ are provided in case (2) and (3), respectively, \myname{} has a noticeable increase in background rejection. This improvement is elucidated in Figure \ref{fig:Scores} where the dashed line represents case (1) and the solid lines represent case (2). Here one can see a dramatic improvement in the lowest bin, background-like, and highest bin, signal-like which proves the the predicted $E_{frac}$ and $M_{frac}$ of the model can be used directly to perform event level classification.

\section{Conclusion}\hfill

We suggest a novel model, \myname{}, to address the effect of the pileup interactions on physics studies at current and future LHC conditions. The model makes use of a stack of transformer encoders with self- and cross-attention using track and jet parameters in the event. The proposed architecture allows the model to learn kinematic correlations arising from physics processes which enables \myname{} to directly predict the fractions of jet energy and jet mass due to pileup in order to recover physically significant observables. The model was trained and tested using simulated di-Higgs datasets. It was shown that the model is capable of recovering the value and resolution of the Higgs boson mass in high-pileup conditions, where initially the Higgs boson mass peak is not possible to observe. \myname{} provides a computationally efficient algorithm to model pileup at the event level and scales well with the expected increase of the pileup level at the LHC and HL-LHC. The OSU high-energy physics group, a member of the ATLAS Collaboration at the LHC, plans to integrate \myname{} into ATLAS software and evaluate its performance with real LHC data.
%
%
%
\bibliographystyle{splncs04}
\bibliography{PileupPaper}
\end{document}